\documentclass[trackchanges]{aastex701}

\begin{document}
\title{MATCH: Performance Characterization and Scientific Observations of the Wuhan University 1-m Optical Telescope}

\author[0009-0007-6828-3931]{Sai-En Xu}
\affiliation{Department of Astronomy, School of Physics and Technology, Wuhan University, Wuhan 430072, People’s Republic of China}
\email{saienxu@whu.edu.cn}

\correspondingauthor{Bei You, Zong-Hong Zhu}
\author[0000-0002-8231-063X]{Bei You}
\affiliation{Department of Astronomy, School of Physics and Technology, Wuhan University, Wuhan 430072, People’s Republic of China}
\email{youbei@whu.edu.cn}

\author[0009-0007-7292-8392]{Han He}
\affiliation{Department of Astronomy, School of Physics and Technology, Wuhan University, Wuhan 430072, People’s Republic of China}
\email{hehan123@whu.edu.cn}

\author{Shuai-Kang Yang}
\affiliation{Department of Astronomy, School of Physics and Technology, Wuhan University, Wuhan 430072, People’s Republic of China}
\email{yangsk@whu.edu.cn}

\author[0000-0001-7350-8380]{Xiao Fan}
\affiliation{Department of Astronomy, School of Physics and Technology, Wuhan University, Wuhan 430072, People’s Republic of China}
\email{whufanxiao@whu.edu.cn}

\author[]{Yi-Fei Qu}
\affiliation{Department of Astronomy, School of Physics and Technology, Wuhan University, Wuhan 430072, People’s Republic of China}
\email{2024300001012@whu.edu.cn}

\author[0009-0002-6492-2920]{Rui-Xiang Hu}
\affiliation{Department of Astronomy, School of Physics and Technology, Wuhan University, Wuhan 430072, People’s Republic of China}
\email{huruixiang@whu.edu.cn}

\author[0009-0009-7686-7798]{Hui-Bo Fei}
\affiliation{Department of Astronomy, School of Physics and Technology, Wuhan University, Wuhan 430072, People’s Republic of China}
\email{2023300002037@whu.edu.cn}

\author[0009-0005-8989-0680]{Bo-An Chen}
\affiliation{Department of Astronomy, School of Physics and Technology, Wuhan University, Wuhan 430072, People’s Republic of China}
\email{chenboan@whu.edu.cn}

\author[]{Zong-Hong Zhu}
\affiliation{Department of Astronomy, School of Physics and Technology, Wuhan University, Wuhan 430072, People’s Republic of China}
\email{zhuzh@whu.edu.cn}

\author[]{Zheng-Yang Li}
\affiliation{Nanjing Institute of Astronomical Optics \& Technology, Chinese Academy of Sciences, Nanjing 210042, People’s Republic of China}
\email{}

\author[]{Xiao-Yan Li}
\affiliation{Nanjing Institute of Astronomical Optics \& Technology, Chinese Academy of Sciences, Nanjing 210042, People’s Republic of China}
\email{}

\author[]{Zi-Jian Han}
\affiliation{Nanjing Institute of Astronomical Optics \& Technology, Chinese Academy of Sciences, Nanjing 210042, People’s Republic of China}
\email{}

\author[]{Jia-Nan Cong}
\affiliation{Nanjing Institute of Astronomical Optics \& Technology, Chinese Academy of Sciences, Nanjing 210042, People’s Republic of China}
\email{}

\author[]{Chao Chen}
\affiliation{Nanjing Institute of Astronomical Optics \& Technology, Chinese Academy of Sciences, Nanjing 210042, People’s Republic of China}
\email{}

\author[]{Jia-Li Chen}
\affiliation{Nanjing Institute of Astronomical Optics \& Technology, Chinese Academy of Sciences, Nanjing 210042, People’s Republic of China}
\email{}

\author[]{Kai-Wen Zheng}
\affiliation{Nanjing Institute of Astronomical Optics \& Technology, Chinese Academy of Sciences, Nanjing 210042, People’s Republic of China}
\email{}

\author[]{Yi-Qiao Yang}
\affiliation{Nanjing Institute of Astronomical Optics \& Technology, Chinese Academy of Sciences, Nanjing 210042, People’s Republic of China}
\email{}

\author[]{Qing-Shan Li}
\affiliation{Astronomical Instruments Co. Ltd., Tianjin, People’s Republic of China}
\email{}

\author[]{Zhen-Guang Sun}
\affiliation{Astronomical Instruments Co. Ltd., Tianjin, People’s Republic of China}
\email{}

\author[]{Tong Zhou}
\affiliation{Nanjing Institute of Astronomical Optics \& Technology, Chinese Academy of Sciences, Nanjing 210042, People’s Republic of China}
\email{}

\author[]{Kai Zhang}
\affiliation{Nanjing Institute of Astronomical Optics \& Technology, Chinese Academy of Sciences, Nanjing 210042, People’s Republic of China}
\email{}

\author[]{Jiajia Wu}
\affiliation{Nanjing Institute of Astronomical Optics \& Technology, Chinese Academy of Sciences, Nanjing 210042, People’s Republic of China}
\email{}

\author[]{Xuhang Yin}
\affiliation{Nanjing Institute of Astronomical Optics \& Technology, Chinese Academy of Sciences, Nanjing 210042, People’s Republic of China}
\email{}

\author[]{Liang Yuan}
\affiliation{Nanjing Institute of Astronomical Optics \& Technology, Chinese Academy of Sciences, Nanjing 210042, People’s Republic of China}
\email{}



\begin{abstract}

The Multi-mode Autonomous Terminal for Compatible Hybrid-optics (MATCH), operated by Wuhan University and located at Lenghu, is a 1-m Ritchey--Chr\'etien telescope equipped with imaging and spectroscopic instruments. MATCH has been in trial operation since October 2025. In this work, we present the telescope and instrument configuration, characterize its performance during the first year of operation, and describe the integrated observing workflow developed for routine and time-domain observations, including the observatory control system, observation scheduler, and automated data reduction pipelines. 
Under typical observing conditions, the imaging system reaches 60-s-equivalent $5\sigma$ limiting magnitudes of approximately 18--20 mag across the $ugri$ bands, while the spectroscopic system reaches approximately 15--16~mag in the B and G channels in a 30-minute integration at the effective spectral resolution of the instrument. In the best-quality imaging observations, the $g$- and $r$-band depths approach 21~mag. Together with flexible scheduling and Target-of-Opportunity capability, MATCH provides photometric and spectroscopic follow-up and long-term monitoring of transient and variable sources.

\end{abstract}




\section{Introduction}
\label{sect:intro}

The rapid development of wide-field surveys, including the All-Sky Automated Survey for Supernovae (ASAS-SN) \citep{2014ApJ...788...48S}, the Asteroid Terrestrial-impact Last Alert System (ATLAS) \citep{2018PASP..130f4505T}, the Zwicky Transient Facility (ZTF) \citep{2019PASP..131a8002B}, the Vera C. Rubin Observatory's Legacy Survey of Space and Time (LSST) \citep{2019ApJ...873..111I}, and the Wide Field Survey Telescope (WFST) \citep{2023SCPMA..6609512W}, together with space-based missions such as the Einstein Probe (EP) \citep{Yuan2022}, has greatly expanded the scope of time-domain astronomy. These facilities enable systematic monitoring of the variable sky over a broad range of timescales, providing both the discovery of transient events and the characterization of persistent variable sources. Many time-domain science cases, however, require observations beyond those provided by survey programs themselves, including rapid follow-up of newly discovered events, high-cadence observations during selected phases of variability, multi-band photometric and spectroscopic characterization, and long-term monitoring. Small- and medium-aperture telescopes are particularly well suited to these tasks because they can provide flexible scheduling, relatively abundant observing time, and dedicated observations of individual targets. 

The Lenghu site on the Tibetan Plateau has been identified as a high-quality site for optical astronomy, with favorable seeing conditions, a high fraction of photometric nights, and a geographic location well suited to time-domain observations \citep{2021Natur.596..353D}. A number of astronomical facilities have subsequently been deployed or planned at the site. In particular, the 2.5\,m WFST has been in operation since 2023 and conducts wide-field time-domain surveys of the northern sky \citep{2023SCPMA..6609512W}. Alongside such survey facilities, smaller dedicated telescopes can provide complementary capabilities through flexible target-oriented observations, rapid photometric and spectroscopic follow-up, and sustained monitoring over long time baselines.

The Multi-mode Autonomous Terminal for Compatible Hybrid-optics (MATCH) was installed at the Lenghu site in August 2025 and has been in trial operation since October 2025. Equipped with both imaging and spectroscopic instruments, MATCH was developed as a flexible 1-m facility for photometric and spectroscopic follow-up of transient and variable sources, as well as long-term monitoring programs. In this work, we describe the telescope and instrument configuration, characterize its performance during the first year of operation, and present the observing system and workflow developed for routine and time-domain observations.

The remainder of this paper is organized as follows. Section~\ref{sec:facility} describes the telescope and instrument configuration. Section~\ref{sec:ocs} presents the observatory control system, which supports automated imaging observations and semi-automated spectroscopic observations. The data reduction pipelines are introduced in Section~\ref{sec:data_reduction}, and the observational performance of the facility is characterized in Section~\ref{sec:performance}. Section~\ref{sect:ObsScheduler} presents the scientific goals, observing strategy, scheduling system, and operational workflow. The main results are summarized in the final section.

\section{Facility}
\label{sec:facility}

\subsection{Optical Design}

MATCH employs a 1-m, $f/7$ Ritchey--Chr\'etien (R--C) optical system with a Cassegrain focus.  A 1000-mm concave hyperboloidal primary mirror and a 380-mm convex hyperboloidal secondary mirror, separated by 1500~mm, provide an effective focal length of 7000~mm. Including the secondary cell and baffles, the effective obscuration diameter is 420~mm. Table~\ref{tab:optical-parameters} summarizes the principal optical parameters.

\begin{figure}[!htbp]
  \centering
  \includegraphics[width=\textwidth,trim=0 145bp 0 0,clip]{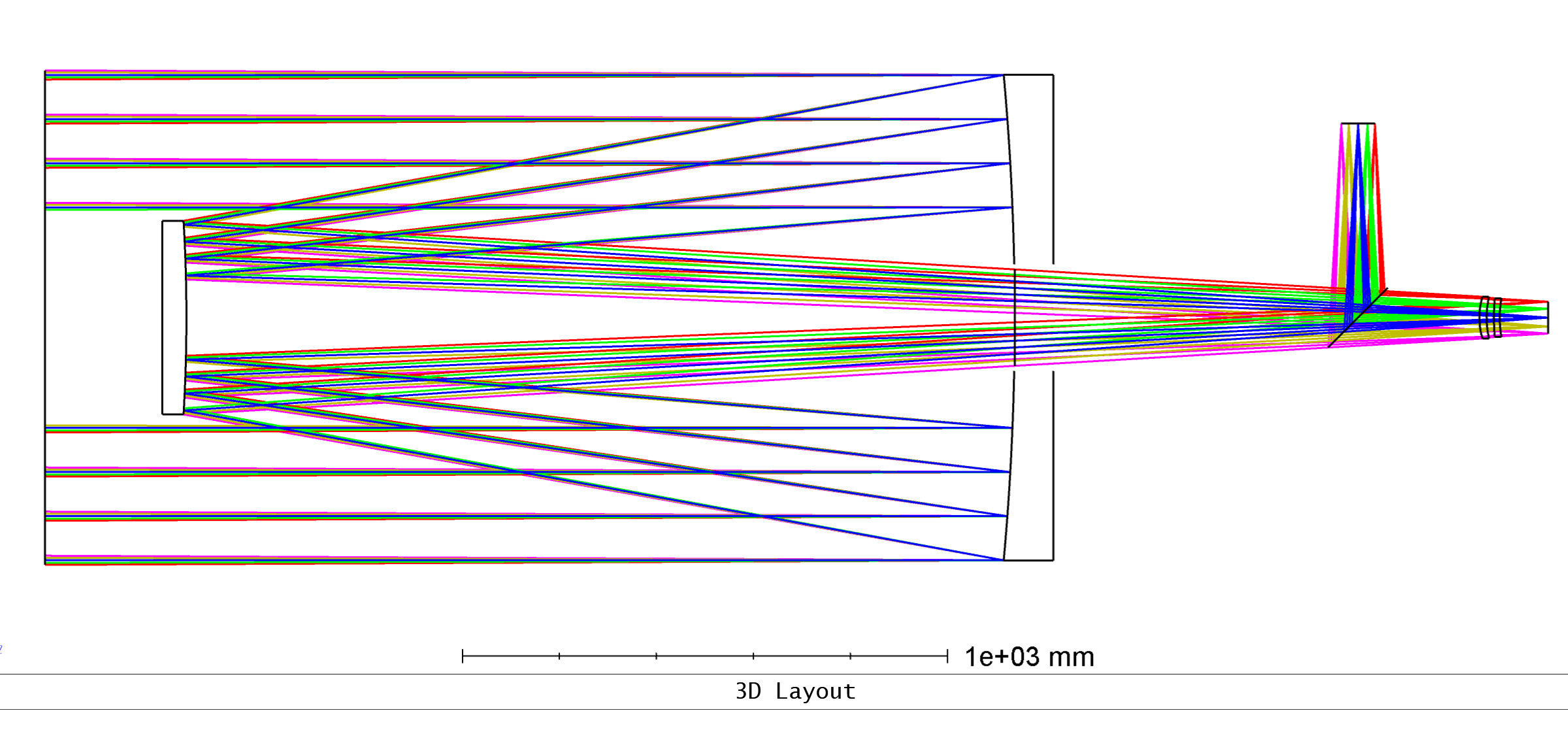}
  \caption{Optical layout of the telescope. The two-element fused-silica corrector is located near the Cassegrain focal plane.}
  \label{fig:optical-layout}
\end{figure}

As shown in Figure~\ref{fig:optical-layout}, the secondary mirror returns the beam through the central aperture of the primary mirror. Near the Cassegrain focus, the beam passes through two spherical fused-silica corrector lenses (L1 and L2), a 3-mm filter, and a 3-mm fused-silica camera window. The corrector suppresses residual astigmatism and field curvature, producing a flat 67.2-mm-diameter corrected image field and expanding the field from approximately $10\arcmin$ without correction to $0.55\degr$ in diameter.
The optical system covers 365--1014~nm and has an image scale of $29.47\arcsec$~mm$^{-1}$. A 3.76-$\mu$m pixel therefore subtends $0.111\arcsec$. The $53.8\times40.4$~mm detector covers $0.44\degr\times0.33\degr$, and its diagonal matches the 67.2-mm corrected image circle.

\begin{deluxetable}{lc}
\tablewidth{0pt}
\tablecaption{Principal optical parameters of MATCH.
\label{tab:optical-parameters}}
\tablehead{
    \colhead{Parameter} & \colhead{Value}
}
\startdata
Entrance pupil diameter & 1000 mm \\
Primary mirror & $R=-4600$ mm, $K=-1.115$ \\
Secondary mirror & $D=380$ mm, $R=-2383$ mm, $K=-5.009$ \\
Effective central obscuration & 420 mm \\
Primary--secondary separation & 1500 mm \\
Effective focal length & 7000 mm \\
Focal ratio & $f/7$ \\
Design wavelength range & 365--1014 nm \\
Corrected field diameter & $0.55\degr$ \\
Corrected image-circle diameter & 67.2 mm \\
Image scale & $29.47\arcsec$ mm$^{-1}$ \\
\enddata
\tablecomments{$R$, $K$, and $D$ denote the radius of curvature, conic constant, and mirror diameter, respectively.}
\end{deluxetable}

\begin{figure}[!htbp]
  \centering
  \includegraphics[width=0.92\textwidth]{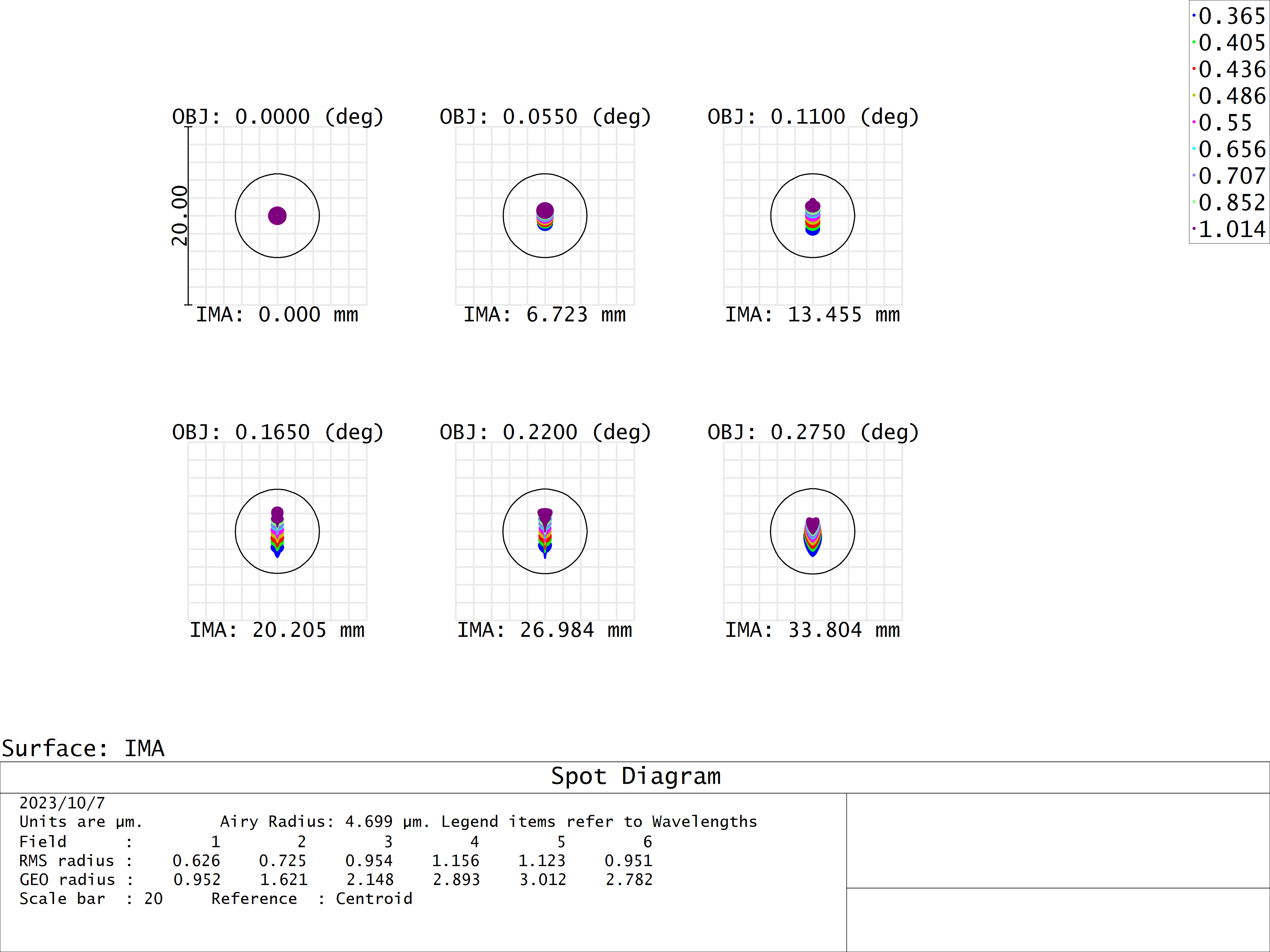}
  \caption{Polychromatic spot diagrams from the center to the edge of the $0.55\degr$-diameter field. Nine wavelengths from 365 to 1014~nm are included; the circles mark the Airy disk.}
  \label{fig:spot-diagram}
\end{figure}

\begin{figure}[!htbp]
  \centering
  \includegraphics[width=0.88\textwidth]{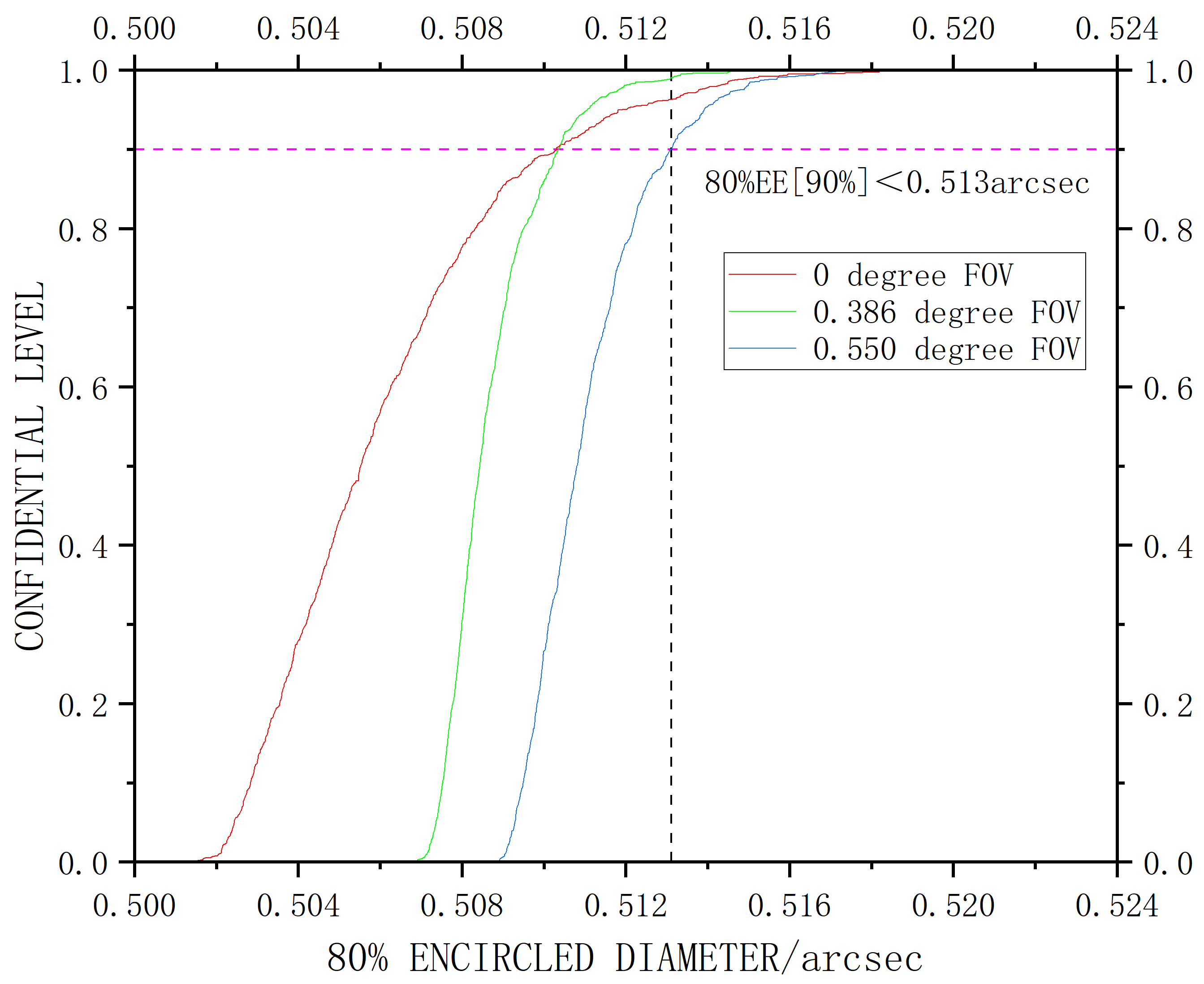}
  \caption{Cumulative EE80 distributions from 1000 Monte Carlo tolerance trials. At the field edge, the 90th-percentile diameter is $0.513\arcsec$ (approximately $0.52\arcsec$).}
  \label{fig:tolerance-montecarlo}
\end{figure}

The nominal image quality was evaluated at nine wavelengths from 365 to 1014~nm. Across the six field positions shown in Figure~\ref{fig:spot-diagram}, the root-mean-square spot radius ranges from 0.626 to 1.156~$\mu$m.
The nominal design requirement is an 80\%-encircled-energy (EE80) diameter below $0.3\arcsec$, and the designed image quality remains diffraction-limited at all evaluated field positions. Engineering feasibility was assessed with 1000 Monte Carlo trials that include fabrication and alignment tolerances. At the field edge, 90\% of the trials remain below $0.52\arcsec$ (Figure~\ref{fig:tolerance-montecarlo}).
Axial refocusing of the secondary mirror and focal-plane assembly compensates the modeled thermal focus shift from $-35$ to $+35\degr$C.

\subsection{Mechanical Structure and Mount Control}

\begin{figure}[htbp]
  \centering
  \includegraphics[width=0.5\textwidth]{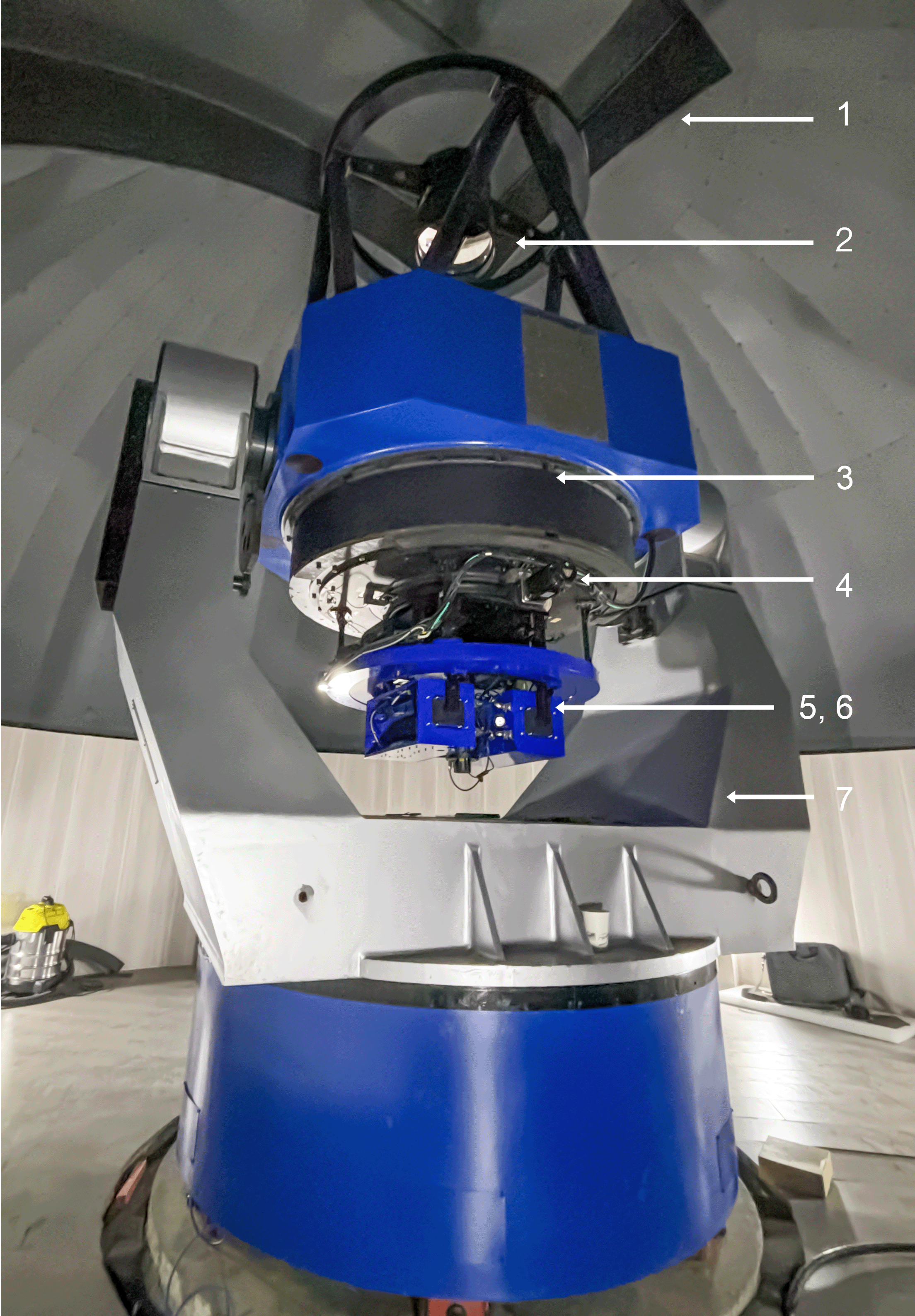}
  \caption{Perspective view of the telescope and its altitude--azimuth mount. 1: The Dome. 2: The secondary mirror. 3: The primary mirror. 4: The derotator. 5, 6: Photometric and spectroscopic instruments. 7: The altitude-azimuth mount.}
  \label{fig:mount-overview}
\end{figure}

MATCH is installed on a direct-drive altitude--azimuth mount, as shown in Figure~\ref{fig:mount-overview}. A Serrurier truss connects the primary- and secondary-mirror assemblies and is designed to preserve their relative alignment as the telescope elevation changes. The primary mirror rests on an 18-point axial whiffletree with inner and outer rings of 6 and 12 supports. Fourteen lever-counterweight supports, divided equally above and below the mirror, carry the lateral load. The optimized support limits the calculated primary-mirror deformation to less than 15~nm RMS in the horizontal orientation and 6~nm RMS in the vertical orientation. The secondary mirror uses a three-point flexure support with a central Invar locator. A five-axis mechanism provides alignment adjustment, while its motorized axial stage provides focus control. A Heidenhain AT1218 length gauge closes the focus loop over a 12-mm travel, with a 368-nm measurement step and $\pm1~\mu$m positioning accuracy.

Coreless axial-flux torque motors drive both axes directly. The azimuth axis uses a self-aligning rolling-bearing long shaft, whereas each altitude axis uses a back-to-back pair of angular-contact bearings. The mount covers $-270\degr$ to $+270\degr$ in azimuth and $0\degr$ to $+90\degr$ in altitude, with a maximum slew rate of $2\degr$~s$^{-1}$ on either axis. Absolute Renishaw steel-tape encoders on 300-mm rings close the position loops; their specified resolution and accuracy are below $0.001\arcsec$ and $1.9\arcsec$, respectively.

A Sidereal Technology ForceTwo controller implements nested current, velocity, and position loops, supplemented by velocity and acceleration feedforward. Its ASCOM interface links the mount to the observatory control system without moving the low-level servo loops to the supervisory layer. At the Cassegrain focus, a field derotator compensates the rotation introduced by the altitude--azimuth mount. The derotator combines a stepper motor, a backlash-compensated dual-gear transmission, and a Heidenhain ECA~4402 absolute angular encoder in a closed position loop.

\subsection{Instruments}

MATCH is equipped with two instruments at the Cassegrain focus: a broadband imaging system for photometric observations and a long-slit spectrograph. The two instruments are mounted on separate ports and can be selected according to the observing program. Their main specifications are summarized in Tables~\ref{tab:photometry_instrument} and \ref{tab:spectrograph_instrument}.

The imaging system uses a QHYCCD QHY411 camera equipped with a back-illuminated Sony IMX411 CMOS sensor. The $14304\times10748$ array has a pixel size of 3.76~$\mu$m, corresponding to a sampling of $0.111\arcsec$~pixel$^{-1}$ at the $f/7$ Cassegrain focus. The resulting field of view is approximately $0.44\degr\times0.33\degr$. Broadband $u$, $g$, $r$, and $i$ filters are available for routine photometric observations.

The spectroscopic instrument is a long-slit spectrograph equipped with a ZWO ASI6200 camera based on the back-illuminated Sony IMX455 CMOS sensor. The detector contains $9576\times6388$ pixels with a pixel size of 3.76~$\mu$m. The spectrograph provides wavelength coverage through three spectral channels and is primarily intended for classification and follow-up spectroscopy of transient and variable sources. The resolving power ranges from approximately $R=1000$ to $R=3000$, depending on the adopted slit width.

\begin{deluxetable}{lc}
\tablewidth{0pt}
\tablecaption{Main specifications of the imaging instrument.
\label{tab:photometry_instrument}}
\tablehead{
    \colhead{Parameter} & \colhead{Value}
}
\startdata
Science camera & QHY411 (IMX411) \\
Field of view & $0.44\degr \times 0.33\degr$ \\
Filters & $u$, $g$, $r$, $i$ \\
\enddata
\end{deluxetable}

\begin{deluxetable}{lcc}
\tablewidth{0pt}
\tablecaption{Main specifications of the spectroscopic instrument.
\label{tab:spectrograph_instrument}}
\tablehead{
    \colhead{Parameter} & \colhead{Configuration} & \colhead{Value}
}
\startdata
Science camera & --- & ASI6200 (IMX455) \\
Guide camera & --- & ASI220 \\
Spectral coverage & Blue & 390--516 nm \\
                  & Green & 506--680 nm \\
                  & Red & 670--906 nm \\
Slit length & --- & $8\arcmin$ \\
Resolving power & $1\arcsec$ slit & $R=3000$ \\
                & $2\arcsec$ slit & $R=2000$ \\
                & $3\arcsec$ slit & $R=1000$ \\
Guide-camera field of view & --- & $3.84\arcmin \times 2.16\arcmin$ \\
\enddata
\end{deluxetable}

\subsection{Detector Characterization}
\label{sec:detectors}

Both MATCH instruments employ back-illuminated scientific CMOS detectors. Although the general characteristics of the Sony IMX411 and IMX455 detector families have been investigated by \citet{2023PASP..135e5001A}, quantities such as electronic gain and readout noise depend on the adopted camera configuration. We therefore characterized both detectors independently using the settings adopted for routine observations. 

Following the manufacturer's recommendations, the QHY411 is operated in readout \textit{mode~\#4}, with gain = 0 and offset = 50. This mode provides relatively low readout noise while retaining a large dynamic range. For the ASI6200, the readout noise decreases substantially near gain = 100; we therefore adopt gain = 100 and offset = 50 for routine spectroscopic observations.

The detector characteristics were measured using calibration data obtained with the same configurations as those adopted for routine observations. Readout noise was measured from pairs of bias frames acquired over several months of regular operation. Following the standard two-bias method, the readout noise was calculated as
\begin{equation}
\sigma_{\rm RN}=\frac{g\,\sigma(B_1-B_2)}{\sqrt{2}},
\end{equation}
where $B_1$ and $B_2$ are two bias frames, $g$ is the measured electronic gain in $e^{-}\,\mathrm{ADU}^{-1}$, and $\sigma(B_1-B_2)$ is the standard deviation of the difference image in ADU. Differencing removes the fixed bias structure, while the factor of $\sqrt{2}$ accounts for the independent readout-noise contributions from the two frames. The bias level itself remained highly stable over this period: the median bias of the spectroscopic camera showed no measurable variation at the ADU level, while that of the imaging camera varied by only about 1~ADU. The resulting readout noise measurements were therefore representative of the normal operating conditions of both detectors.

Dark current was measured from dedicated dark frame sequences obtained at different detector temperatures and exposure times. The electronic gain of the imaging detector was determined from twilight sky-flat exposures spanning a range of signal levels, whereas that of the spectroscopic detector was derived from flat-field lamp exposures acquired at different signal levels. The resulting electronic gain, readout noise, and dark-current measurements are summarized in Table~\ref{tab:detector}.

\begin{deluxetable}{lcc}
\tablewidth{0pt}
\tablecaption{Specifications and measured characteristics of the CMOS detectors used by MATCH.
\label{tab:detector}}
\tablehead{
    \colhead{Parameter} &
    \colhead{Imaging camera} &
    \colhead{Spectroscopic camera}
}
\startdata
Camera & QHYCCD QHY411 & ZWO ASI6200 \\
Sensor & Sony IMX411 & Sony IMX455 \\
Array size (pixels) & $14304 \times 10748$ & $9576 \times 6388$ \\
Pixel size ($\mu$m) & 3.76 & 3.76 \\
\cutinhead{Operational configuration}
Readout mode & Mode~\#4 & --- \\
Gain setting & 0 & 100 \\
Offset setting & 50 & 50 \\
\cutinhead{Measured characteristics}
Electronic gain ($e^{-}\,\mathrm{ADU}^{-1}$) & 1.007 & 0.257 \\
Readout noise ($e^{-}$) & 3.07 & 1.10 \\
Dark current at $-20^\circ$C
($e^{-}\,\mathrm{pixel}^{-1}\,\mathrm{ks}^{-1}$)
& 2.052 & 0.0267 \\
Dark current at $-10^\circ$C
($e^{-}\,\mathrm{pixel}^{-1}\,\mathrm{ks}^{-1}$)
& 2.401 & 0.1717 \\
\enddata
\tablecomments{The electronic gain, readout noise, and dark current were measured in this work using the detector configurations adopted for routine observations.}
\end{deluxetable}



\section{Observatory Control System}
\label{sec:ocs}

To facilitate the efficient execution of scientific observing programs, we developed the MATCH Observatory Control System (MATCH OCS) in Python to coordinate telescope operation and data acquisition through a unified graphical interface. Imaging observations can be executed automatically from predefined observation plans, whereas spectroscopic observations are conducted semi-automatically. For each scheduled target, the workflow coordinator interprets the target coordinates and observing mode, checks altitude and telescope-motion constraints, configures the required instrument, controls telescope slewing and focusing, and executes the requested exposures.

To maintain image quality as the observing conditions and telescope orientation vary, the MATCH OCS uses an empirical model to predict the approximate focal position. Empirical focus models are commonly used to account for repeatable focus variations with temperature and telescope altitude or elevation \citep{1996PASP..108..220H,10.1117/1.OE.52.5.053604}. In robotic operation, such predictions can provide an initial position for a subsequent image-based autofocus sequence \citep{2010AdAst2010E..32B}, while accumulated focus measurements can in turn be used to refine the empirical relation \citep{2010AdAst2010E..57S}. For MATCH, the predicted focal position is described by
\begin{equation}
f(T,h)=A\sin h+BT+C_{\rm mode},
\end{equation}
where $T$ is the ambient temperature, $h$ is the telescope altitude, and $C_{\rm mode}$ is a constant offset that differs between the imaging and spectroscopic optical paths. The predicted position is used as the center of a local autofocus scan. At each scan position, stellar sources are detected using SEP \citep{2016JOSS....1...58B,1996A&AS..117..393B}, and their half-flux radii are measured. A hyperbolic V-curve is then fitted to the measured image sizes to determine the best-focus position.

The PyQt6-based graphical interface accepts operator commands, observation plans, and system configuration parameters, and reports observing progress and system status. Within the OCS core, device management handles hardware connections, serializes access to shared devices, and maintains device status, while image analysis and data recording are kept separate from low-level hardware communication. Acquired images are stored in FITS format together with metadata describing the observing mode, instrument configuration, target coordinates, exposure timing, and detector temperatures, providing the information required by the subsequent data-reduction pipeline.

\begin{figure}
    \centering
    \includegraphics[width=\linewidth]{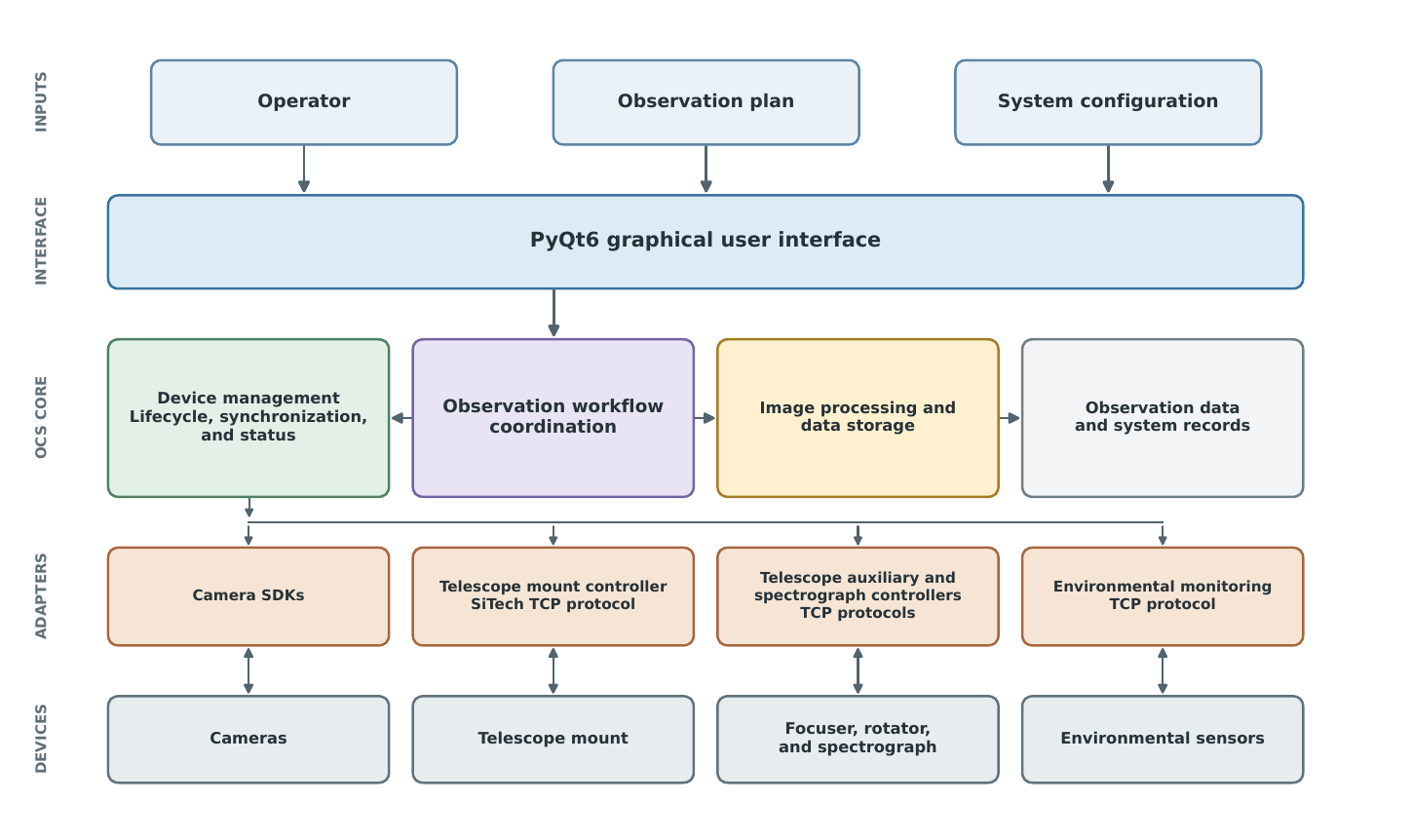}
    \caption{Overall architecture of the MATCH Observatory Control System. Operator input, observation plans, and system configuration are provided through a PyQt6 graphical interface. The OCS core coordinates target-constraint checks, telescope pointing, model-guided focusing, exposure execution, and observation records, while hardware-specific communication is handled through dedicated device adapters.}
    \label{fig:OCS}
\end{figure}

Hardware-specific communication is isolated in an adapter layer. Camera access is implemented through the corresponding vendor software development kits, while the telescope mount, auxiliary telescope systems, spectrograph mechanisms, and environmental sensors are controlled through their respective TCP-based interfaces. As illustrated in Figure~\ref{fig:OCS}, this layered architecture connects the scientific observing workflow to observatory hardware while keeping device-specific commands separate from the high-level observing logic.

\section{Data reduction pipelines}
\label{sec:data_reduction}

\subsection{Photometric Reduction}

Automated photometric reduction begins with the transfer of raw imaging frames from the MATCH archive at Lenghu to the local processing server. Science frames are calibrated using master bias, dark, and flat-field frames with \texttt{ccdproc} \citep{matt_craig_2025_16755473}. Sources are extracted with SExtractor \citep{1996A&AS..117..393B}, and astrometric calibration is performed locally with Astrometry.net \citep{2010AJ....139.1782L} using offline index files, yielding the World Coordinate System (WCS) solution for each frame.  The resulting solutions are validated against the expected target coordinates, and only frames with reliable astrometry and target coverage are retained for subsequent analysis. For routine observations of isolated point sources, photometric measurements and calibration are performed with AutoPhOT \citep{2022A&A...667A..62B}. 

For sources embedded in spatially structured extended emission, such as galactic nuclei, changes in seeing can alter the contribution of the surrounding host galaxy within a fixed aperture and introduce systematic variations into conventional aperture or PSF photometry. We therefore developed an additional reduction procedure based on PSF homogenization and reference-image differential photometry. Depending on the target and data quality, the images can either be convolved to a common PSF before fixed-aperture photometry or processed using difference imaging. The latter follows the general approach of PSF-matched image subtraction \citep{1998ApJ...503..325A,2000A&AS..144..363A} and is implemented using HOTPANTS \citep{2015ascl.soft04004B}.

For difference imaging, the calibrated frames are resampled onto a common astrometric grid, and sources and background properties are characterized using SEP \citep{2016JOSS....1...58B,1996A&AS..117..393B}. A reference image is selected using image-quality metrics including the FWHM, PSF shape, background level, number of detected sources, and usable image area. Residual astrometric offsets relative to the reference are corrected before subtraction. HOTPANTS determines the convolution kernel required to match the point spread functions and photometric scaling of the reference and science images before producing the difference images. The target region is excluded from the kernel determination to prevent intrinsic source variability from affecting the PSF-matching solution.

Photometry is subsequently performed at the target position on the difference images. The target flux in the reference image is measured independently, and the total flux at each epoch is reconstructed as $F_i=F_{\rm ref}+\Delta F_i$, where $\Delta F_i$ is the differential flux measured from the corresponding difference image. The absolute photometric scale is determined from catalog stars in the reference image using robust outlier rejection. Difference images are subjected to automated quality checks based on the background noise, residuals of field stars, large-scale background structure, and the number of significant residual detections; epochs that fail these checks are excluded from the final light curve. Photometric uncertainties include the uncertainties of the reference and differential flux measurements together with that of the photometric zero point.


\subsection{Spectroscopic Reduction}

Spectroscopic data can be reduced interactively using the standard IRAF long-slit tasks through \texttt{PyRAF} \citep{1986SPIE..627..733T,1993ASPC...52..173T,2012ascl.soft07011S}, providing flexibility for manual inspection and specialized reductions. For routine and time-domain observations, however, rapid assessment of the acquired spectra is desirable, particularly for Target-of-Opportunity (ToO) programs. We therefore developed an automated long-slit reduction pipeline based on \texttt{PyLongslit} \citep{2025JOSS...10.9264V} and adapted it to the MATCH spectrograph.

The pipeline uses FITS metadata and instrument-specific configuration files to associate calibration frames, identify the B, G, and R spectral channels, and select the appropriate detector regions. Detector calibration, two-dimensional wavelength calibration, sky modeling and subtraction, one-dimensional spectral extraction, and, where suitable spectrophotometric standards are available, atmospheric-extinction correction and flux calibration are carried out automatically. 

An additional validation step was developed to assess the wavelength solution directly from the science exposures. The primary wavelength calibration is derived from manually reviewed FeAr reference solutions for each spectral channel. Night-sky emission features detected in individual science frames are then compared with their expected wavelengths and, when sufficiently significant, are used to verify and refine the wavelength zero point. This provides an independent check of wavelength-calibration stability during unattended processing.

\begin{figure}
    \centering
    \includegraphics[width=0.9\linewidth]{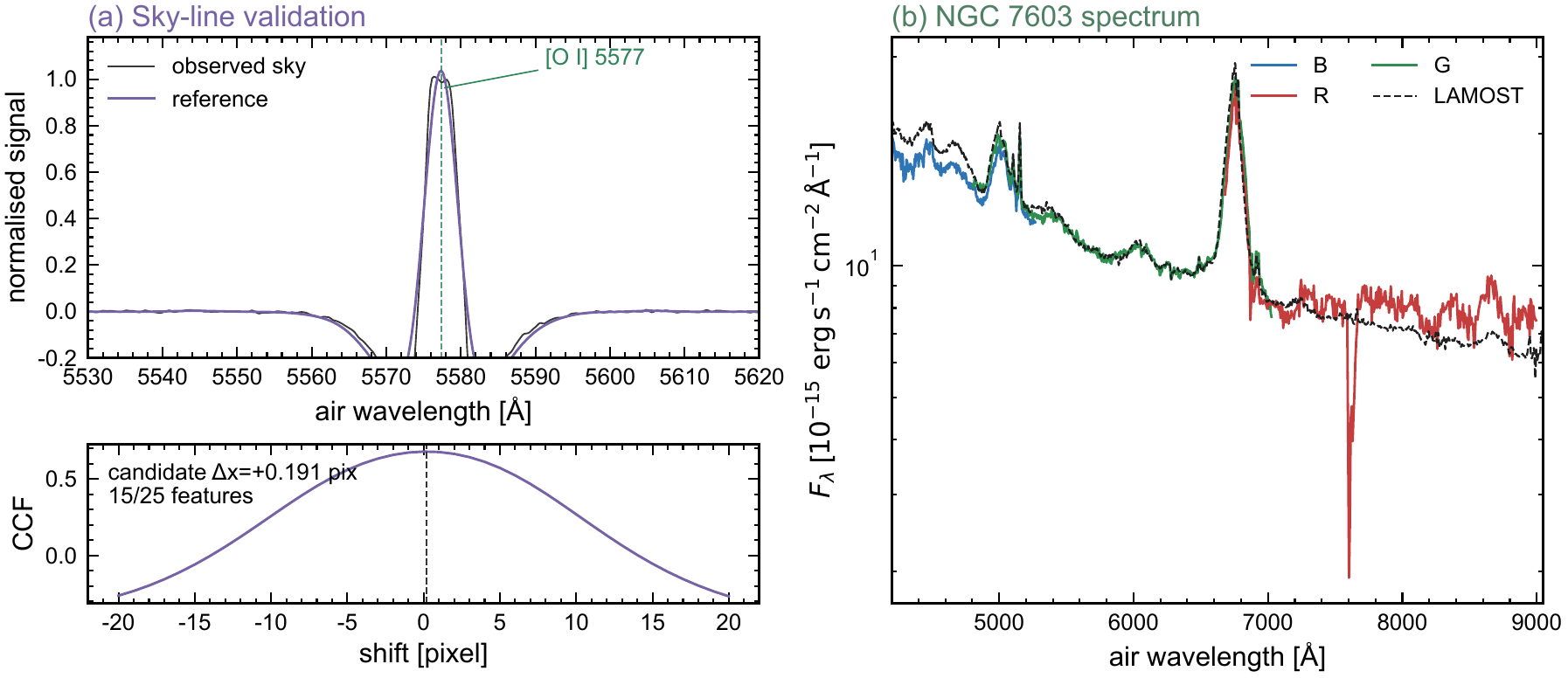}
    \caption{Examples of products from the automated spectroscopic reduction pipeline. Left: wavelength-zero-point validation using the [O~I] night-sky emission line near 5577~\AA. Right: reduced MATCH spectrum of NGC~7603 compared with the corresponding LAMOST spectrum after normalization over a common continuum interval. }
    \label{fig:spec_pipe}
\end{figure}

Figure~\ref{fig:spec_pipe} shows representative products of the automated reduction. The night-sky-line comparison illustrates the wavelength-zero-point validation, while the reduced spectrum of NGC~7603 is compared with an archival LAMOST spectrum as an external consistency check. Differences in the detailed continuum shape are expected because the observations were obtained with different instruments, apertures, and observing conditions, and may also reflect residual uncertainties in the relative flux calibration and wavelength-dependent slit losses.

\section{Performance}
\label{sec:performance}

\subsection{Photometric Performance}

To characterize the photometric performance of MATCH, we analyzed imaging observations obtained during the first year of operation in the $u$, $g$, $r$, and $i$ bands. We focus on the photometric zero point, limiting magnitude, and color term, which characterize the system throughput, sensitivity, and transformation between the MATCH and reference photometric systems, respectively. Each image was processed with \textsc{AutoPhOT} \citep{2022A&A...667A..62B}. For each exposure, instrumental magnitudes were calibrated against catalogue photometry according to
\begin{equation}
m_{{\rm cat},j}=m_{{\rm inst},ij}+ZP_i+c\,C_j,
\end{equation}
where $ZP_i$ is the zero point of exposure $i$, $c$ is the color coefficient for the corresponding filter, and $C_j$ is the catalogue color index of star $j$. The zero point was allowed to vary between exposures to account for changes in atmospheric transparency and observing conditions.

\begin{figure}
    \centering
    \includegraphics[width=\linewidth]{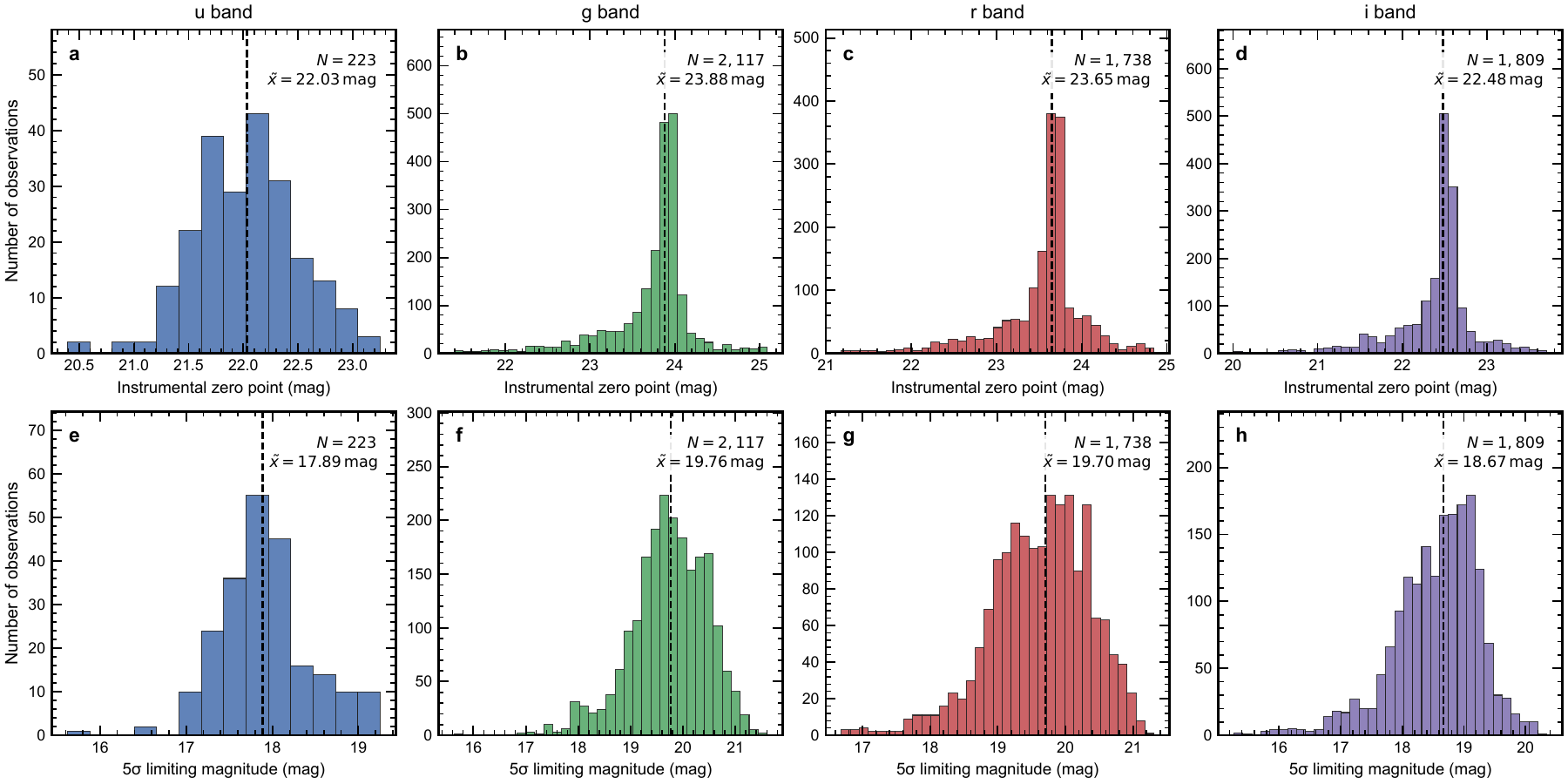}
    \caption{Distributions of the long-term photometric performance in the $u$, $g$, $r$, and $i$ bands. The four columns correspond to the $u$, $g$, $r$, and $i$ filters from left to right, respectively. The upper and lower rows show the photometric zero point and the 60-s-equivalent $5\sigma$ limiting magnitude, respectively. The vertical dashed line marks the median value. The value of $N$ in each panel denotes the number of images included in the corresponding distribution.}
    \label{fig:photo_performance}
\end{figure}

The long-term distributions of the zero point and $5\sigma$ limiting magnitude are shown in Figure~\ref{fig:photo_performance}, and the corresponding median values and central 68\% intervals are summarized in Table~\ref{tab:color_terms}. The limiting magnitude of each exposure was estimated from the empirical relation between calibrated magnitude and signal-to-noise ratio and evaluated at $\mathrm{SNR}=5$. For comparison between observations with different exposure times, the limiting magnitudes listed in Table~\ref{tab:color_terms} were scaled to an equivalent exposure time of 60~s assuming background-limited scaling. The median $5\sigma$ depths are approximately 17.9, 19.8, 19.7, and 18.7~mag in the $u$, $g$, $r$, and $i$ bands, respectively. In the best-quality exposures, the corresponding 60-s-equivalent limiting magnitudes approach 19~mag in $u$, 20~mag in $i$, and 21~mag in both $g$ and $r$.

Color terms were derived from the accumulated AutoPhOT calibration products using Pan-STARRS DR2 reference stars. Calibration stars were restricted to $14\leq m\leq17$ and screened for photometric quality and blending. For each exposure, the stellar zero points were fitted as a linear function of catalogue color after robust outlier rejection. Only frames containing a sufficient number of calibration stars over an adequate color baseline were retained. The accepted per-exposure slopes were combined using a block bootstrap resampled by observing night, thereby accounting for correlations among exposures obtained during the same night. Pan-STARRS colors $(g-r)_{\rm PS1}$, $(r-i)_{\rm PS1}$, and $(i-z)_{\rm PS1}$ were adopted for the $g$, $r$, and $i$ bands, respectively. No corresponding Pan-STARRS color term was derived for the $u$ band. The resulting coefficients are listed in Table~\ref{tab:color_terms}. The color-term uncertainties are the 16th--84th percentile range of 500 block-bootstrap realizations, with resampling performed at the level of individual observing nights to preserve intra-night correlated systematics.

\begin{deluxetable*}{lcccccc}
\tablewidth{0pt}
\tablecaption{Summary of the first-year photometric performance of MATCH}
\label{tab:color_terms}
\tablehead{
\colhead{Filter} &
\colhead{$N_{\rm img}$} &
\colhead{Zero point} &
\colhead{$5\sigma$ limiting magnitude} &
\colhead{$N_{\rm img,CT}$} &
\colhead{Color index} &
\colhead{Color term}
}
\startdata
$u$ & 223 & $22.03^{+0.44}_{-0.44}$ & $17.89^{+0.51}_{-0.45}$ & \nodata & \nodata & \nodata \\
$g$ & 2117 & $23.88^{+0.14}_{-0.51}$ & $19.76^{+0.71}_{-0.68}$ & 508 & $(g-r)_{\rm PS1}$ & $-0.044^{+0.002}_{-0.001}$ \\
$r$ & 1738 & $23.65^{+0.19}_{-0.52}$ & $19.70^{+0.65}_{-0.75}$ & 552 & $(r-i)_{\rm PS1}$ & $-0.020^{+0.001}_{-0.001}$ \\
$i$ & 1809 & $22.48^{+0.16}_{-0.46}$ & $18.67^{+0.51}_{-0.73}$ & 386 & $(i-z)_{\rm PS1}$ & $-0.144^{+0.003}_{-0.003}$ \\
\enddata
\tablecomments{Zero points and limiting magnitudes are given in magnitudes. The quoted values are medians, with the lower and upper uncertainties corresponding to the 16th and 84th percentiles, respectively. Limiting magnitudes are normalized to an exposure time of 60~s and correspond to $\mathrm{SNR}=5$. $N_{\rm img}$ denotes the number of images used to characterize the zero-point and limiting-magnitude distributions, while $N_{\rm img,CT}$ denotes the independently screened frames used for the color-term analysis. Color terms were measured using calibration stars with $14\leq m\leq17$. No Pan-STARRS color term was derived for the $u$ band.}
\end{deluxetable*}

\subsection{Spectroscopic Performance}

\begin{figure*}
    \centering
    \includegraphics[width=0.9\textwidth]{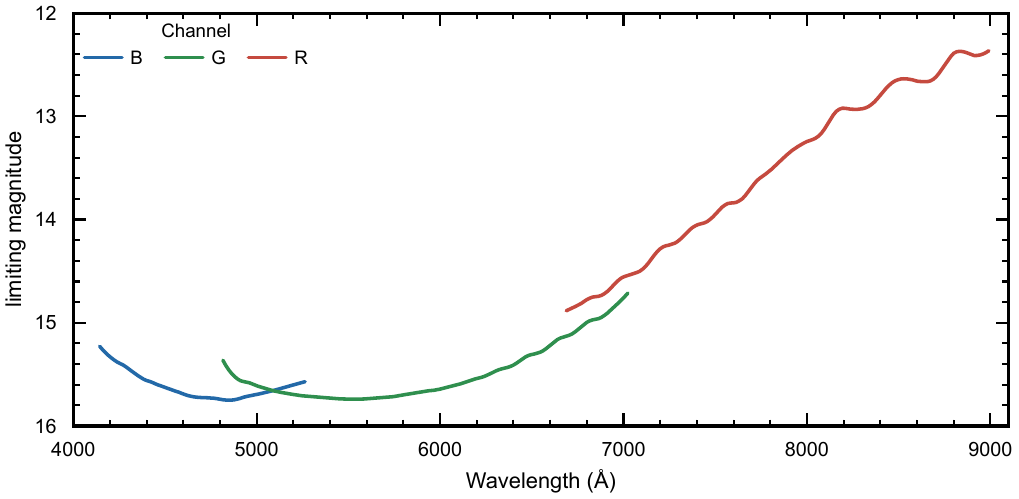}
    \caption{Estimated $5\sigma$ spectroscopic limiting magnitude as a function of wavelength for a 30-minute integration in the B, G, and R channels. The spectra are rebinned to 5~\AA\ intervals, approximately matching one effective spectral resolution element for the adopted configuration ($R\sim1000$), and the limiting magnitudes therefore correspond to an SNR of 5 per resolution element. A representative dark-sky background and detector noise are included in the calculation.}
    \label{fig:Spec_sensitivity}
\end{figure*}

The spectroscopic sensitivity of MATCH was characterized using observations of spectrophotometric standard stars in the B, G, and R channels. After detector calibration and wavelength calibration, the detected source counts were extracted as a function of wavelength and compared with the incident photon flux expected from the calibrated spectral energy distributions of the standards. The wavelength-dependent end-to-end throughput was estimated as
\begin{equation}
T(\lambda)=\frac{N_{\rm det}(\lambda)}{N_{\rm inc}(\lambda)},
\end{equation}
where $N_{\rm det}(\lambda)$ is the number of detected photoelectrons and $N_{\rm inc}(\lambda)$ is the number of photons incident on the telescope aperture over the same wavelength interval and exposure time. The measured throughput was then combined with the representative sky background, detector dark current, and readout noise to estimate the signal-to-noise ratio per resolution element for sources of different magnitudes. As the spectrograph is oversampled along the dispersion direction, the spectra were rebinned to 5~\AA\ intervals, approximately matching one effective spectral resolution element for the adopted configuration ($R\sim1000$). Figure~\ref{fig:Spec_sensitivity} shows the resulting $5\sigma$ limiting magnitude for a 30-minute exposure.

The relatively long $8\arcmin$ slit of the MATCH spectrograph also enables spatially resolved observations of nearby extended sources. By stepping the telescope in the direction perpendicular to the slit, a sequence of long-slit spectra can be combined to reconstruct a three-dimensional $(x,y,\lambda)$ data cube. Although this observing mode is less efficient than a dedicated integral-field spectrograph, the long spatial coverage of the slit makes it potentially useful for mapping nebulae and nearby galaxies over angular scales of several arcminutes. 

\begin{figure*}
    \centering
    \includegraphics[width=0.85\textwidth]{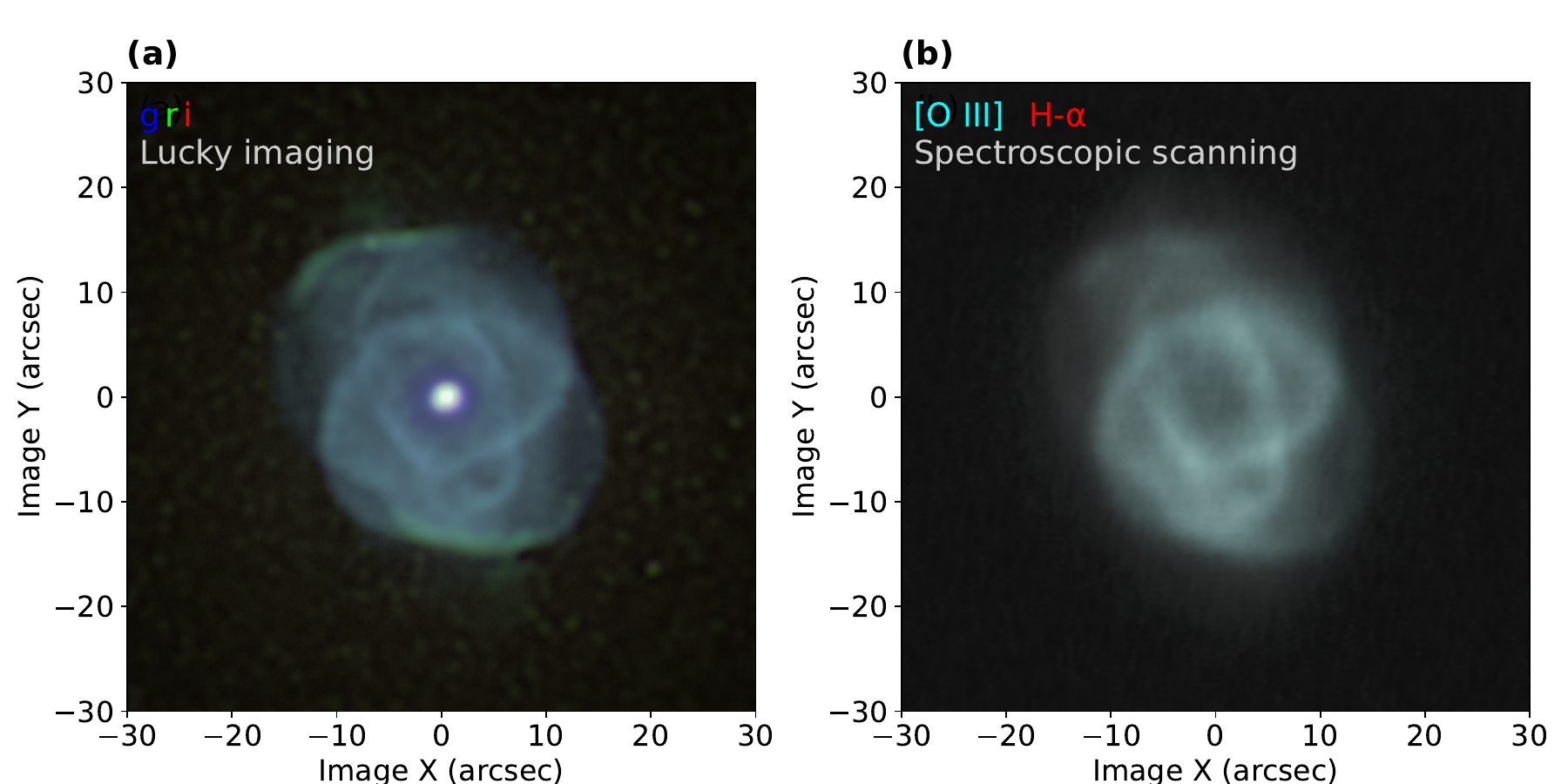}
    \caption{Broadband and narrowband images of NGC 6543. (a) Broadband image of NGC~6543 obtained by lucky imaging with the imaging camera in the $g$, $r$, and $i$ bands. (b) Narrowband image reconstructed from the three-dimensional spectra around [O~III] and H$\alpha$. The scan step is $1\arcsec$, with an integration time of 2~s per step and a total integration time of 80~s. }
    \label{fig:NGC6543}
\end{figure*}

We tested this observing mode on NGC~6543. The slit was stepped across the target in $1\arcsec$ increments with an integration time of 2~s at each position. The resulting spectra were assembled into a three-dimensional data cube, from which narrowband images around the [O~III] and H$\alpha$ emission lines were extracted. Figure~\ref{fig:NGC6543} compares the reconstructed emission-line image with a broadband $gri$ image of the same target obtained with the imaging camera using lucky imaging.

\section{Science Programs and Operations}
\label{sect:ObsScheduler}

MATCH is designed as a dedicated follow-up facility for time-domain astronomy, providing flexible photometric and spectroscopic observations of variable and transient sources identified by wide-field surveys. Its combination of multi-band imaging, long-slit spectroscopy, flexible scheduling, and long-term observing availability makes it well suited to both sustained monitoring campaigns and rapid follow-up of newly discovered transients. The scientific program focuses on binary systems, active galactic nuclei, tidal disruption events, and quasi-periodic eruptions.

\subsection{Binary Systems}
Binary systems remain fundamental laboratories for testing stellar structure, evolution, and interaction physics. The telescope's scheduling flexibility enables sustained photometric monitoring of eclipsing binaries, cataclysmic variables, and compact binary systems (including white dwarfs and X-ray binaries) over timescales spanning individual orbital periods to multi-year baselines. Scientific objectives include: precise determination of orbital periods and their long-term evolution through eclipse timing \citep{Burdge2019ApJ...886L..12B,Burdge2019Natur.571..528B,Burdge2020ApJ...905L...7B,Burdge2022Natur.605...41B,Burdge2022Natur.610..467B,Burdge2023ApJ...953L...1B,Burdge2024Natur.635..316B,Jain2011MNRAS.413....2J};
monitoring of outburst and quiescence cycles in cataclysmic variables and low-mass X-ray binaries to constrain accretion disk instability models \citep{Lasota2001NewAR..45..449L,Dubus2018A&A...617A..26D,Hameury2020AdSpR..66.1004H,You2023Sci...381..961Y,He2026ApJ..1006..177H};
and spectroscopic follow-up of newly identified eclipsing or ellipsoidal-variable binaries \citep{Prsa2011AJ....141...83P,Keller2022MNRAS.509.4171K,Torres2025A&A...698A.173T}. The ability to interleave long-term cadence observations with target-of-opportunity (ToO) triggers is particularly valuable for capturing binaries during rare outburst or superoutburst episodes.

\subsection{Tidal Disruption Events and Quasi-Periodic Eruptions}

Tidal disruption events (TDEs) occur when a star wanders close to a massive black hole and is torn apart, powering luminous flares that probe otherwise quiescent black holes \citep{Gezari2021ARAA..59...21G,vanVelzen2020SSRv..216..124V,Kara2025ARAA..63..379K}.
Wide-field optical surveys now dominate TDE discoveries,
delivering rising light curves and spectroscopic classifications \citep{Gezari2012Natur.485..217G,vanVelzen2020SSRv..216..124V,Hammerstein2023ApJ...942....9H},
while flexible 1-m-class facilities remain essential for timely spectroscopic follow-up and high-cadence multi-band photometric coverage.

The telescope will contribute primarily through sustained monitoring and ToO spectroscopy of survey-selected TDE transients. Scientific objectives include:
(i) characterizing repeating and partial TDEs, in which bound stars produce recurrent optical flares on timescales from months to decades
\citep{Payne2021ApJ...910..125P,Lin2024ApJ...971L..26L,Wevers2023ApJ...942L..33W,Sun2024A&A...692A262S};
(ii) photometric follow-up of extreme TDE phenomena,
including fast-rising or off-nuclear TDEs as IMBH candidates \citep{Angus2022NatAs...6.1452A,Yao2025ApJ...985L..48Y},
TDE-like nuclear transients with pre-existing AGN accretion \citep{Blanchard2017ApJ...843..106B,Petrushevska2023A&A...669A140P},
relativistic (jetted) TDEs \citep{Bloom2011Sci...333..203B,Cenko2012ApJ...753...77C,Andreoni2022Natur.612..430A};
and (iii) high-cadence spectroscopic follow-up of optically bright TDEs to track emission-line profiles, evolution, and continuum--line lags \citep{Leloudas2019ApJ...887..218L,Hung2019ApJ...883...44H,Hung2020ApJ...903...31H,Charalampopoulos2022AandA.659A..16C,Holoien2016MNRAS.455.2918H,Nicholl2019MNRAS.488.1878N}.

There is a growing observational link between TDEs and X-ray quasi-periodic eruptions (QPEs): QPEs have now been discovered following the optical TDEs and TDE-like nuclear transients \citep{Nicholl2024Natur.634..804N,Chakraborty2025ApJ...983L..39C}. Spectroscopy of QPE hosts and their preceding nuclear activities constrains the galaxy environments and nuclear states of these systems \citep{2024A&A...688A.157S,Wevers2022AandA.659A...2W,Wevers2024ApJ...970L..23W,Wevers2024ApJ...969L..17W}.
Also, recent detection of a delayed ultraviolet counterpart to Ansky's X-ray QPEs \citep{Guo2026ApJL.1000L..57G} raises the possibility of correlated optical variability on hour-to-day timescales.
The telescope will explore these science opportunities.

\subsection{Active Galactic Nuclei}
AGN variability across the optical, ultraviolet, and X-ray bands provides a powerful probe of the structure and physics of the accretion disk and the surrounding broad-line region (BLR). MATCH can contribute to AGN studies through high-cadence reverberation-mapping campaigns and long-term photometric and spectroscopic monitoring. Key scientific goals include continuum reverberation mapping to constrain the characteristic size and structure of AGN accretion disks, together with reverberation studies of low-mass AGNs and intermediate-mass black-hole (IMBH) candidates \citep{Rafter2011ApJ...741...66R,Guo2022ApJ...940...20G,Wang2023ApJ...948L..23W,Zuo2024ApJ...974..288Z,2024ApJ...976..176F,2026ApJ...998..311L,2026ApJ...997..326F,2025ApJ...986..137Z}; long-term optical monitoring to estimate black-hole masses and investigate AGN variability through variability-based scaling relations and stochastic models such as the damped random walk \citep{Kelly2009ApJ...698..895K,MacLeod2010ApJ...721.1014M,Kasliwal2015MNRAS.451.4328K,Burke2021Sci...373..789B,Yu2022ApJ...936..132Y}; and long-term photometric and spectroscopic monitoring of changing-look AGNs (CLAGNs), which exhibit dramatic changes in continuum emission and broad emission lines on timescales of months to years \citep{LaMassa2015ApJ...800..144L,MacLeod2019ApJ...874....8M,Ricci2023NatAs...7.1282R,He2026arXiv260606802H}. These studies benefit particularly from dense and sustained temporal sampling over long baselines, which is essential for characterizing AGN variability across a wide range of timescales.

\subsection{Observation Scheduling}

\begin{figure}
    \centering
    \includegraphics[width=0.9\linewidth]{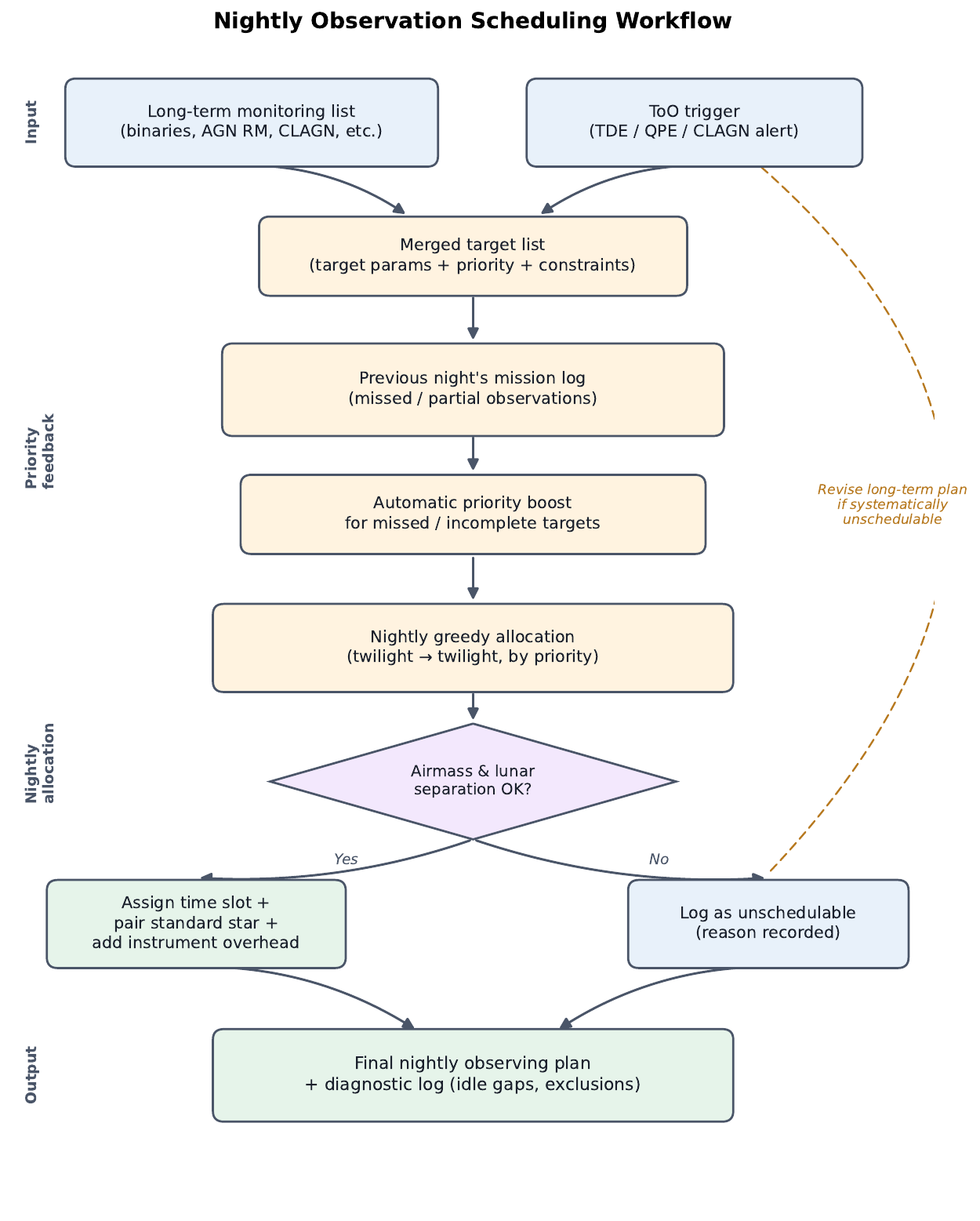}
    \caption{Workflow of the automated nightly scheduling pipeline.}
    \label{fig:scheduler}
\end{figure}

Ground-based follow-up requires balancing reliable long-term monitoring with rapid response to unpredictable alerts. Our scheduler merges both into one nightly plan: targets carry priority, visibility constraints, and instrument setup; the previous night's log automatically boosts priorities for missed or incomplete observations, preventing cadence gaps in long-baseline campaigns; a greedy, priority-ordered algorithm then allocates time slots across the night subject to airmass and lunar-separation limits, pairing standard stars with science targets and logging any unschedulable cases. ToO alerts simply enter the same queue with elevated priority, so rapid follow-up and sustained monitoring share one consistent workflow suited to a single-telescope facility with limited nightly observing time. In Figure \ref{fig:scheduler} we present the workflow of the scheduler.

\subsection{Closed-loop Operational Workflow}

The OCS, scheduler, reduction pipelines, and web interface are integrated into a closed operational workflow connecting proposal submission, observation planning, telescope operation, data reduction, and scientific feedback. Approved observing programs are entered into the scheduling system, which generates nightly plans according to visibility, priority, cadence, and instrumental requirements. The plans are then executed through the OCS.

Raw data acquired at Lenghu are transferred automatically to the data-processing server, where the photometric and spectroscopic pipelines generate reduced and quick-look products. These products are made available through the web interface, allowing observers to assess data quality and scientific progress and, when necessary, modify the priorities or cadence of subsequent observations. This feedback loop enables both long-term monitoring programs and rapid follow-up observations to be adjusted on a night-to-night basis.

\section{Conclusion}
\label{sec:conclusion}

MATCH has been in trial operation since October 2025, providing nearly one year of routine imaging and spectroscopic observations for evaluating the performance and stability of the facility. Using these data, we characterized the photometric performance and spectroscopic sensitivity of the telescope and its instruments. Under typical observing conditions, the imaging system reaches a $5\sigma$ limiting magnitude of approximately 20~mag in the $g$ band for a 60~s exposure, while the spectroscopic system reaches approximately 15.5~mag in the B and G channels for a 1800~s integration at the effective spectral resolution of the instrument.

To support efficient time-domain observations, we also developed an integrated operational workflow linking proposal submission, observation scheduling, telescope control, automated data reduction, and quick-look data products. The scheduling system accommodates both long-term monitoring programs and Target-of-Opportunity observations, allowing regular cadence observations and rapid transient follow-up to be handled within a common operational framework.

The first year of operation demonstrates that MATCH can provide flexible photometric and spectroscopic follow-up of transient and variable sources, together with sustained monitoring over long temporal baselines. Continued operation will be used to refine the automation and calibration procedures, improve the long-term characterization of the facility, and expand its contribution to time-domain observing programs.

\begin{acknowledgments}
This work was funded by “the Fundamental Research Funds for the Central Universities”; the National Natural Science Foundation of China (NSFC) under No. 12322307, 12273026, and 12361131579; Xiaomi Foundation / Xiaomi Young Talents Program. The data analysis presented in this paper was performed using computational resources provided by the Supercomputing Center of Wuhan University. We thank Cheng Cheng, Hai-Cheng Feng and Sha-Sha Li for helpful comments and suggestions regarding both the manuscript and the operation of the telescope.
\end{acknowledgments}

\begin{contribution}

B.Y. initiated and led the MATCH project as the Principal Investigator, defined the scientific program, supervised the work, secured funding, and contributed to the writing of the manuscript. Z.-H.Z. co-supervised the project and contributed to the manuscript. S.-E.X. conducted detector characterization, developed the OCS, the basic imaging data-reduction pipeline, and the spectroscopic reduction pipeline, and wrote the corresponding sections of the manuscript. S.-K.Y. and R.-X.H. developed the imaging data-reduction pipeline and wrote the corresponding section. H.-B.F. conducted the optical photometric performance analysis and wrote the corresponding section. Y.-F.Q. conducted the spectral performance analysis and wrote the corresponding section. H.H. designed the nightly observation scheduling workflow and wrote the corresponding section. X.F. and B.-A.C. contributed to the project and wrote the sections corresponding to their respective work. Z.-Y.L., X.-Y.L., Z.-J.H., J.-N.C., C.C., J.-L.C., K.-W.Z., Y.-Q.Y., T.Z., K.Z., J.W., X.Y., and L.Y. were responsible for the design, development, integration, and testing of the telescope, and drafted the instrument description section. Q.-S.L. and Z.-G.S. contributed to the telescope fabrication and technical support.


\end{contribution}

%
\facility{MATCH (QHY411 imaging camera, long-slit spectrograph with ASI6200 camera)}

\software{
Astropy \citep{2013A&A...558A..33A,2018AJ....156..123A,2022ApJ...935..167A},
ccdproc \citep{matt_craig_2025_16755473},
SExtractor \citep{1996A&AS..117..393B},
Astrometry.net \citep{2010AJ....139.1782L},
AutoPhOT \citep{2022A&A...667A..62B},
SEP \citep{2016JOSS....1...58B},
HOTPANTS \citep{2015ascl.soft04004B},
IRAF \citep{1986SPIE..627..733T,1993ASPC...52..173T},
PyRAF \citep{2012ascl.soft07011S},
PyLongslit \citep{2025JOSS...10.9264V}
}

\bibliography{PASPsample701}{}
\bibliographystyle{aasjournalv7}



\end{document}